# Superconductivity of Bulk Abnormal Magic-stoichiometric Na$_3$Cl Salt Crystals at Normal Pressure

Shuqiang He, [†a] Yi-Feng Zheng, [†bc] Guosheng Shi, [d] Yi-Jie Xiang, [ce] Meihui Xiao, [a] Qituan Zhang, [a] Yue-Yu Zhang, [*bc] Haiping Fang [*af]

[a] School of Physics, East China University of Science and Technology, Shanghai 200237, China
[b] Wenzhou Institute, University of Chinese Academy of Sciences, Wenzhou 325001, China
[c] University of Chinese Academy of Sciences, Beijing 100049, China
[d] Shanghai Applied Radiation Institute, State Key Lab. Advanced Special Steel, Shanghai University, Shanghai 200444, China
[e] Shanghai Institute of Applied Physics, Chinese Academy of Science, Shanghai 201800, China
[f] School of Physics, Zhejiang University, Hangzhou 310027, China.

[*] Corresponding authors. E-mails: fanghaiping@ecust.edu.cn; zhangyy@wiucas.ac.cn
[†] Equally contributed to this work.

**Abstract**
**The identification of new materials with superconducting properties is the pursuit in the realm of superconductivity research. Here, excitedly, we show that the simplest salt daily used can be made a superconductor at normal pressure only by adjusting its stoichiometry of Na and Cl as Na$_3$Cl at normal pressure based on first-principles calculations. This *bulk* stable abnormal Na-Cl stoichiometric crystal of 3:1, the first 'magic' ratio, includes metallic (Na) atoms in the core as well as hybridization of ionic and metallic bonding, facilitating the electron-phonon-coupling for superconductivity with a critical temperature Tc of 0.13 K. The flat bands and van Hove singularities near the Fermi level produce large densities of states, similar to H$_3$S and LaH$_{10}$, which is beneficial for the emergence of superconductivity. The crystal composed of with abnormal Na-Cl magic stoichiometry is a precisely tunable, purely sodium and chloride-based, three-dimensional bulk superconductor, which is therefore an ideal material for designing and understanding abnormal stoichiometric crystals. The methodology of constructing this bulk abnormal crystal may be general to almost all elements, which could lead to insights into the physics of other conventional superconductors and even high-critical-temperature superconductors.**

Superconductors have always obtained great attention[1,2]. However, up to now, it remains an open question of whether superconductivity is possessed by most and even all elements or some given elements of a specific characteristic. This is the key to overcome challenges in the field of superconductivity research to identify new kind of materials and search room-temperature superconducting materials at normal pressure.

Historically, bulk crystals composed of main group and transition metal elements, exhibiting unconventional stoichiometry, have predominantly emerged under ultra-high pressure conditions[3-6], yielding remarkable properties such as near-room-temperature superconductivity[5,7], transparent inorganic electrides[8], and high-energy-density characteristics[9]. However, stabilizing such structures under ambient conditions has long posed a formidable challenge[10-13]. Based on traditional stoichiometry frameworks to find new superconductivity, especially for room-temperature superconducting materials, the research has encountered bottlenecks and restricted our thinking. Exotic stoichiometric bulk crystals at normal pressure have a great potential to disclose exciting properties[14,15].

Here, by only adjusting the stoichiometry of Na and Cl as the magic ratio $Na_3Cl$, the simplest salt daily used can be made a superconductor at normal pressure based on first-principles calculations. This cation-rich abnormal *bulk* $Na_3Cl$ crystal includes metallic (Na) atoms in the core and hybridization of ionic and metallic bonding, facilitating the electron-phonon-coupling for superconductivity, and we show its superconductivity with a critical temperature Tc of 0.13 K. The methodology of this bulk abnormal crystal construction may be general to almost all elements, illustrating that superconductivity maybe possessed by most and even all elements.

**Computational Methods**

We performed the structure searches for the binary $(Na_3Cl)_x$ system (x = 1–5) using the IM$^2$ODE (Inverse Design of Materials by Multi-objective Differential Evolution)[16]. The calculations of electronic properties were carried out by using PAW potentials in the Vienna ab initio simulation package (VASP)[17]. The projector augmented waves (PAW) approach was used to describe the interaction between ions and electrons[18,19], and the exchange-correlation functional was described by adopting the Perdew-Burke-Ernzerhof (PBE) generalized gradient approximation (GGA) [20]. We chose the cutoff energy of 1000 eV and Monkhorst-Pack ***k*** meshes of $2\pi \times 0.03$ Å$^{-1}$ to ensure the enthalpy calculations accurately converge to 1 meV/atom. Lattice-dynamical and superconducting properties were estimated using the Allen-Dynes modified McMillan equation, performed within the Quantum ESPRESSO package [21,22] with ultrasoft pseudopotential adopted. The kinetic energy cutoff was set to 50 Ry. The ***k***-points of 12 × 12 × 1 and ***q***-points of 4 × 4 × 1 were employed.

**Results and Discussion**

The ground state structure of $Na_3Cl$ found by global optimization without external pressure is a six-layer structure with space group $P2_1/m$ (number 11), as depicted in Figure 1a and 1b. There are 2 Cl atoms and 6 Na atoms in the unit cell of Na$_3$Cl_ $P2_1/m$, and from bottom up, it is composed of Na-Cl layer, 4 pure Na layers, and then Na-Cl layer, totally six atomic layers to form this quasi-bulk crystal. To identify different kinds of Na atoms, here we mark Na1 as Na atoms in the Na-Cl layer, Na2 as Na atoms directly touching with the Na-Cl layer, and Na3 as the rest (Figure 1a). The lattice parameters are a=b=3.88 Å, c= 33.39 Å, $\alpha = 90.00°$, $\beta = 88.97°$, and $\gamma = 90.00°$ with atomic positions at Na1, 2e (0.27, 0.75, 0.39), Na2, 2e (0.24, 0.75, 0.54), Na3, 2e (0.22, 0.75, 0.69) and Cl, 2e (0.28, 0.75, 0.30). The bond length of Na3-Cl is 2.74 Å, which is 26% smaller than the Na-Cl bond length in NaCl rock salt phase. It will result in an increase of the Coulombic attraction between Na cations and Cl anions. The bond length in Na layers, Na1-Na1 bond length is 3.74 Å, which is slightly (1%) larger than the Na bond length of 3.72 Å in its metal body center cubic (bcc) phase. The interlayer bond length of Na1-Na2 and Na2-Na3 are 3.64 Å and 3.96 Å, respectively, which are slightly (2%) smaller and 6% larger than that in bcc phase.

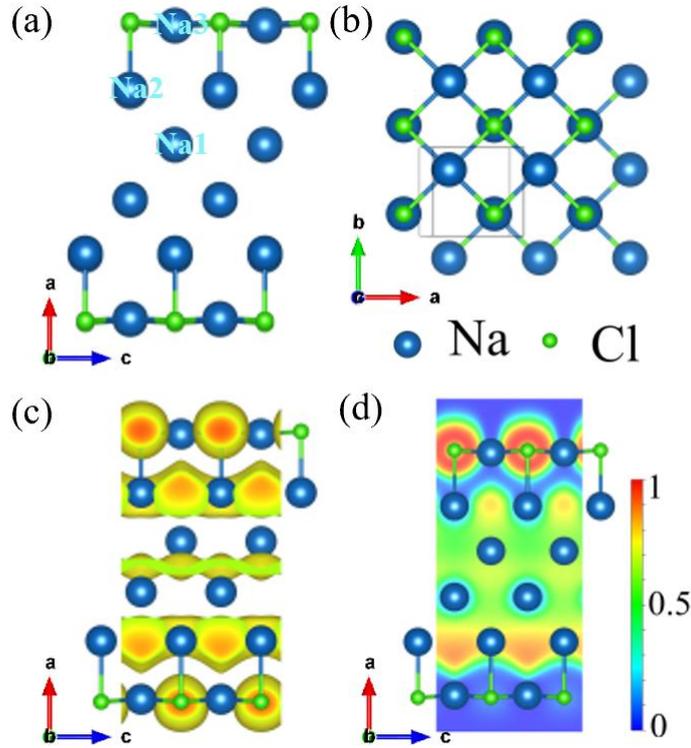

Figure 1 Structure and Electron Locational Fuction (ELF) of quasi-bulk Na$_3$Cl crystal. (a) Side and (b) top view. Na and Cl atoms are presented by blue and green spheres, respectively. (c) ELF with an isosurface value of 0.58 $e/Bohr^3$. (d) Two-dimensional (2D) slide of ELF on (100) plane.

The stability of Na$_3$Cl is originated from both the charger-transfer-induced ionic interaction and metallic bonding interaction like Na layer in the metallic bcc body. In the Na-Cl layer, each Cl gains 0.89 e$^-$ while Na1 losses 0.81 e$^-$, and charge is closely balanced in this layer, like NaCl ionic crystal. Thus, in the Na2 layer, even if Na2 is bonded with Cl, there is minor charge transfer of Na2 loses 0.10 e$^-$. In the Na3 layer, each Na3 atom gains 0.02 e$^-$, similar as the case in metallic bcc Na. If periodic boundary condition is applied to this structure, the bulk NaCl$_3$ is still stable according to both the ionic and metallic interactions, so our result could also be applied in the bulk system. Here for the convenience of calculation, we use the six-layer quasi-bulk phase.

The inner layers of Na show strong metallic properties. The electronic energy band diagrams and density of states diagrams of Na$_3$Cl_ *P2$_1$/m* crystals predicted in this work are shown in Figure 2a. The contribution of Na to the density of states near the Fermi level overwhelms that of Cl. Figure 2b further analyzes the contribution of different Na atoms to the density of states near the Fermi level, and the projected density of states on Na1 and Na2 (inner layers) is much larger than that on Na3 (Na-Cl layer).

The electronic band structure shows three bands crossing the Fermi level ($E_F$) in a parallel manner contributed by Na. These bands feature a simultaneous occurrence of flat bands (Γ point), corresponding to three sharp van Hove singularities at -2.8 eV, -1.9 eV, and -0.7 eV below $E_F$, similar to H$_3$S and LaH$_{10}$[19-22]. These electronic structures have been shown to be favorable for the formation of stable Cooper pairs, resulting in superconductor behavior.

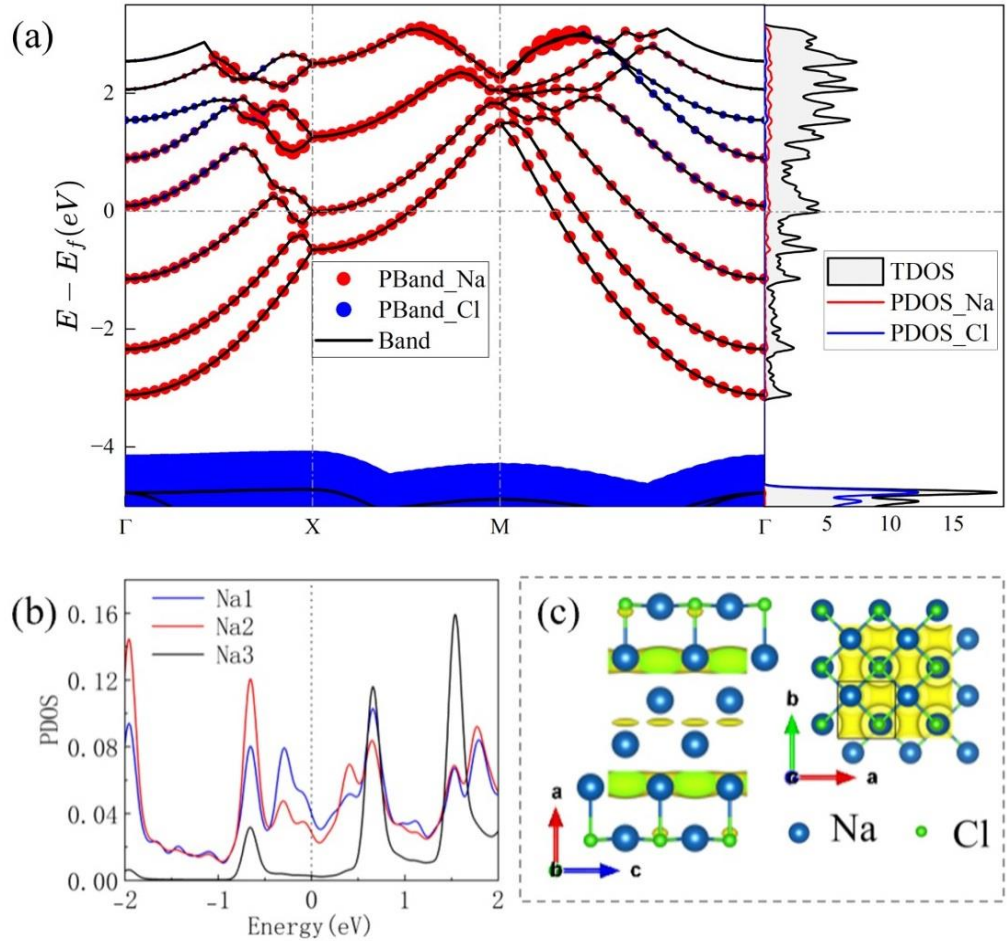

Figure 2 Electronic structure of quasi-bulk Na$_3$Cl crystal. (a) The projected band structure (left panel) and the projected density of state PDOS (right panel) of Na$_3$Cl_*P2$_1$/m*, the green box represents the enlarged view of the density of states near the Fermi energy level. (b)PDOS of ground-state sodium atoms in three different environments near the Fermi level. (c) partial charge density (isosurface value of 0.002e/Bohr3) within 1eV below E$_f$ of Na$_3$Cl_*P2$_1$/m* crystal.

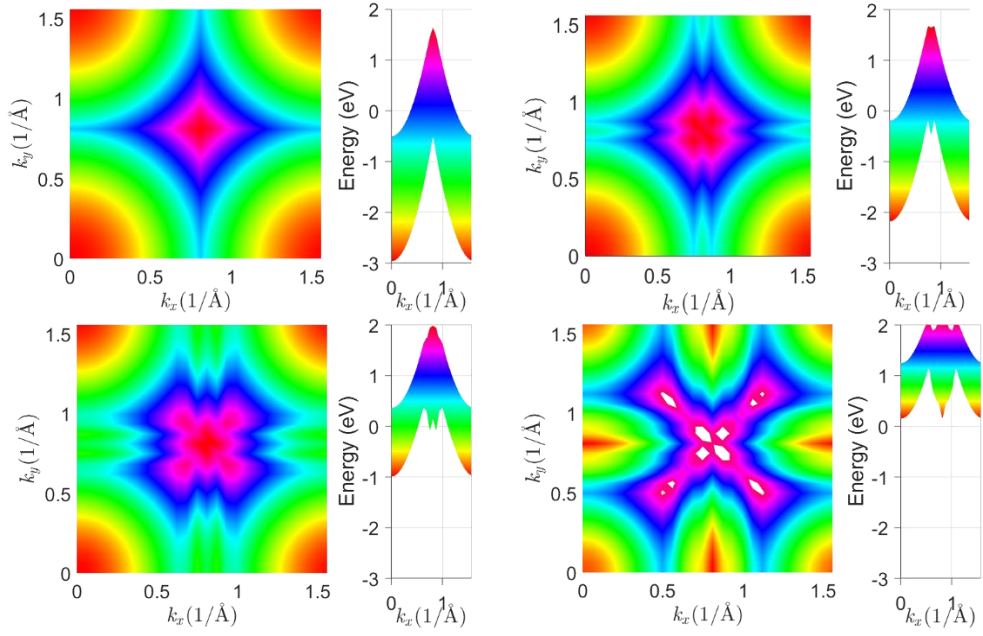

Fig. 3. Three-dimensional visualization of energy bands near the Fermi energy level. There are three energy bands across the Fermi energy level. Additionally, there is an energy band in close proximity to the Fermi energy level. The metallic character of Na$_3$Cl_*P21/m* is attributed to the contribution of the Na atom pair.

The band structures near the Fermi energy level in Figure 3 indicate the possible presence of Fermi surface nesting. The metallic behavior is evident from the bands. We can also find nearly parallel bands and contour cut by the Fermi energy level, which is consistent with the Fermi surface in Figure 4b and presents the possibility of Fermi surface nesting.

According to the Bardeen-Cooper-Schrieffer (BCS) theory, the calculated Electron-Phonon Coupling (EPC) indicates that the superconductivity is due to the soft phonon mode. To investigate the superconducting properties of the quasi-bulk Na$_3$Cl crystals at normal pressure, the Eliashberg spectral function $\alpha^2 F(\omega)$ and the frequency-dependent coupling $\lambda(\omega)$ were calculated. This result is confirmed by the further analysis in Figure.4(a). The coupling constant is calculated by

$$\lambda = \Sigma_{qv}\lambda_{qv} = 2\int_0^\infty d\omega \frac{a^2 F(\omega)}{\omega}$$

We can see that the contributions come from the phonon modes near 2-4 THz. It is worth noting that the optical vibrational modes corresponding to 2 THz exhibit a phonon softening phenomenon, which is reflected by the large projected phonon linewidths and a high peak in the Eliashberg spectral function. Phonon vibrational modes are involved in the EPC in two different frequency ranges: the low-frequency range (< 4 THz), where translational vibrations account for 94% of the total EPC λ; the mid-frequency branch (>4 THz) is mainly associated with Cl atoms, which contribute to 6% of the total λ. Figure 4(b) shows the Fermi surface of bulk $Na_3Cl$ crystals, with a view under the first Brillouin zone on the left and the primitive Brillouin zone view on the right. The Fermi surface has three energy bands that cross it, forming a multilayered columnar nesting in the first Brillouin zone. The colors on the Fermi surface indicate the magnitude of its Fermi velocity, with red indicating a greater Fermi velocity.

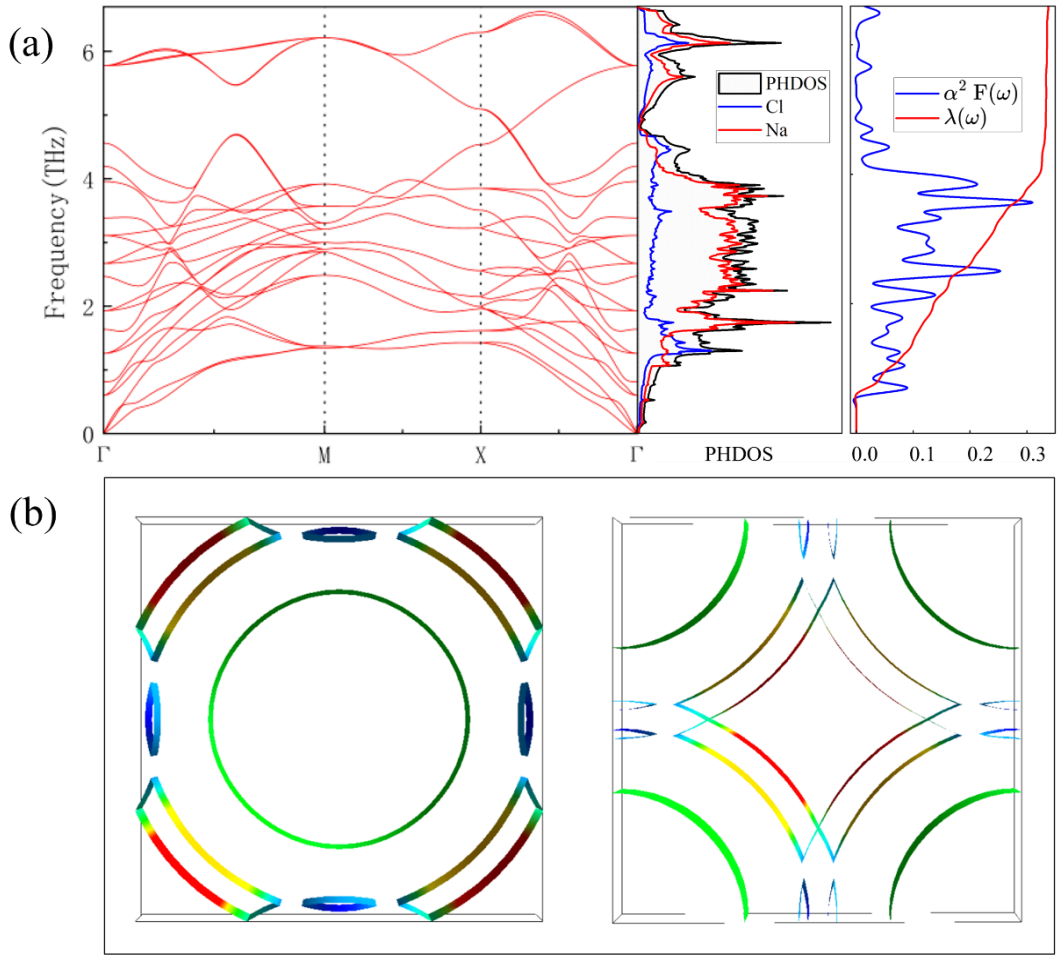

Fig. 4. Calculated phonon properties and Fermi surface of $Na_3Cl$. (a) the phonon dispersion curves, phonon density of states (PHDOS), Eliashberg spectral function $\alpha^2 F(\omega)$ and the accumulated Electron-Phonon Coupling (EPC) constant λ(ω). (b) The Fermi surface of quasi-bulk $Na_3Cl$ crystals, with a view under the first Brillouin zone

on the left and the primitive Brillouin zone view on the right.

The superconducting transition temperature (TC) is estimated using McMillan-Allen-Dynes formula[18]

$$T_C = \frac{\omega_{\log}}{1.2 k_B} \exp\left[\frac{-1.04(1+\lambda)}{\lambda(1-0.62\mu^*)-\mu^*}\right].$$

where $k_B$ is the Boltzmann constant, µ* is the effective screenedCoulomb repulsion constant, typically ~ 0.1, λ is electron-phonon coupling constant, and $\omega_{log}$ is logarithmic average phonon frequency. Phonon softening of acoustic branches is observed around the Γ point. In our quasi-bulk Na3Cl crystals system, λ= 0.38 and $\omega_{log} = 96.63$. Using a Coulomb pseudopotential of $\mu^*$ = 0.1, the calculated *Tc* is 0.13 K.

**Conclusion**
By only adjusting the stoichiometry of Na and Cl as magic ratio Na$_3$Cl, we for the first time found the simplest salt crystals daily used can be made an electron-phonon-coupled superconductor with a $T_c$ of 0.13 K at normal pressure based on first-principles calculations. The key to the superconductivity of this quasi-bulk abnormal magic-stoichiometric Na$_3$Cl crystal is the cation-rich structure including metallic (Na) atoms in the core and hybridization of ionic and metallic bonding, facilitating the electron-phonon-coupling for superconductivity.

It is worth mentioning that the structures and behavior of this quasi-bulk Na$_3$Cl crystal could be extended to bulk crystal. Via applying periodic boundary condition through the z direction, the stability of this bulk phase is confirmed by the local optimization by first-principles calculations. Moreover, the methodology of this bulk abnormal crystal constructing may be general to almost all elements, illustrating that superconductivity may be possessed by most and even all elements. Notably, its bands feature a simultaneous occurrence of flat bands (Γ point), corresponding to three sharp van Hove singularities at -2.8 eV, -1.9 eV, and -0.7 eV below $E_F$, similar to H$_3$S and LaH$_{10}$. In the Na$_3$Cl crystals, the relatively large masses of the Na and Cl ions, which are not able to produce high frequency phonon modes, result into a very small *Tc*. We expect a high value of *Tc* by using lighter atoms, such as Li or F. Thus, there is hope to achieve room temperature superconductivity at normal pressures.

**Acknowledgments**
This work is supported by the Natural Science Foundation of WIUCAS (WIUCASQD2022025, WIUCASQD2023004), National Natural Science Foundation of China (No. 12104452), Shanghai Science and Technology Innovation Action Plan (23JC1401400), and Natural Science Foundation of Wenzhou (L2023005). The 1       Drozdov, A., Eremets, M., Troyan, I., Ksenofontov, V. & Shylin, S. I. Conventional superconductivity at 203 kelvin at high pressures in the sulfur hydride system. *Nature* **525**, 73-76 (2015).
2       Drozdov, A. *et al.* Superconductivity at 250 K in lanthanum hydride under high pressures. *Nature* **569**, 528-531 (2019).
3       Zhang, W. *et al.* Unexpected stable stoichiometries of sodium chlorides. *Science* **342**, 1502-1505 (2013).
4       Zhong, X. *et al.* Tellurium hydrides at high pressures: High-temperature superconductors. *Physical review letters* **116**, 057002 (2016).
5       Duan, D. *et al.* Pressure-induced metallization of dense $(H_2S)_2H_2$ with high-Tc superconductivity. *Scientific reports* **4**, 6968 (2014).
6       Dong, X. *et al.* A stable compound of helium and sodium at high pressure. *Nature Chemistry* **9**, 440-445 (2017).
7       Errea, I. *et al.* Quantum hydrogen-bond symmetrization in the superconducting hydrogen sulfide system. *Nature* **532**, 81-84 (2016).
8       Ma, Y. *et al.* Transparent dense sodium. *Nature* **458**, 182-185 (2009).
9       Christe, K. O., Wilson, W. W., Sheehy, J. A. & Boatz, J. A. $N_5^+$: a novel homoleptic polynitrogen ion as a high energy density material. *Angewandte Chemie International Edition* **38**, 2004-2009 (1999).
10      Miao, M.-s. Caesium in high oxidation states and as a p-block element. *Nature Chemistry* **5**, 846-852 (2013).
11      Botana, J. *et al.* Mercury under pressure acts as a transition metal: calculated from first principles. *Angewandte Chemie International Edition* **54**, 9280-9283 (2015).
12      Yang, G., Wang, Y., Peng, F., Bergara, A. & Ma, Y. Gold as a 6p-element in dense lithium aurides. *Journal of the American Chemical Society* **138**, 4046-4052 (2016).
13      Botana, J. & Miao, M.-S. Pressure-stabilized lithium caesides with caesium anions beyond the −1 state. *Nature communications* **5**, 4861 (2014).
14      Shi, G. *et al.* Two-dimensional Na–Cl crystals of unconventional stoichiometries on graphene surface from dilute solution at ambient conditions. *Nature Chemistry* **10**, 776-779 (2018).
15      Zhang, L. *et al.* Novel 2D CaCl crystals with metallicity, room-temperature ferromagnetism, heterojunction, piezoelectricity-like property and monovalent calcium ions. *National science review* **8**, nwaa274 (2021).
16      Zhang, Y.-Y., Gao, W., Chen, S., Xiang, H. & Gong, X.-G. Inverse design of materials by multi-objective differential evolution. *Computational Materials Science* **98**, 51-55 (2015).
17      Kresse, G. & Furthmuller, J. Efficiency of Ab-initio Total Energy Calculations for Metals and Semiconductors Using a Plane-wave Basis Set. *Comput. Mater. Sci.* **6**, 15-50,

doi:https://doi.org/10.1016/0927-0256(96)00008-0 (1996).

18  Kresse, G. & Joubert, D. From ultrasoft pseudopotentials to the projector augmented-wave method. *Phys. Rev. B* **59**, 1758-1775 (1999).

19  Blöchl, P. E. Projector augmented-wave method. *Phys. Rev. B* **50**, 17953-17979, doi:10.1103/PhysRevB.50.17953 (1994).

20  J. P. Perdew, K. Burke & M. Ernzerhof. Generalized Gradient Approximation Made Simple. *Phys. Rev. Lett.* **77**, 3865, doi:https://doi.org/10.1103/PhysRevLett.77.3865 (1996).

21  Giannozzi, P., S. Baroni, N. Bonini, M. Calandra, R. Car, C. Cavazzoni, D. Ceresoli, G. L. Chiarotti, M. Cococcioni, I. Dabo, A. Dal Corso *et al.* Quantum Espresso: A Modular and Open-source Software Project for Quantum Simulations of Materials. *J. Phys.: Condens. Matter* **21**, 395502, doi:https://doi.org/10.1088/0953-8984/21/39/395502 (2009).

22  Allen, P. B. & Dynes, R. C. Transition temperature of strong-coupled superconductors reanalyzed. *Phys. Rev. B* **12**, 905-922 (1975).